# Electrical detection of surface spin polarization of candidate topological Kondo insulator SmB$_6$

*Jehyun Kim[1\*], Chaun Jang[2\*], Xiangfeng Wang[3], Johnpierre Paglione[3], Seokmin Hong[2], and Dohun Kim[1]*

[1]Department of Physics and Astronomy, and Institute of Applied Physics, Seoul National University, Seoul 08826, Korea

[2]Center for Spintronics, Korea Institute of Science and Technology, Seoul 02792, Korea

[3]Center for Nanophysics and Advanced Materials, Department of Physics, University of Maryland, College Park, MD 20742-4111, USA

[\*]*These authors contributed to this work.*

Correspondence: Dohun Kim, Department of Physics and Astronomy, and Institute of Applied Physics, Seoul National University, Seoul 08826, Korea

E-mail: dohunkim@snu.ac.kr

Running title: Surface spin polarization of SmB$_6$

ABSTRACT: The Kondo insulator compound SmB$_6$ has emerged as a strong candidate for the realization of a topologically nontrivial state in a strongly correlated system, a topological Kondo insulator, which can be a novel platform for investigating the interplay between nontrivial topology and emergent correlation-driven phenomena in solid state systems. Electronic transport measurements on this material, however, so far showed only the robust surface-dominated charge conduction at low temperatures, lacking evidence of its connection to the topological nature by showing, for example, spin polarization due to spin–momentum locking. Here, we find evidence for surface state spin polarization by electrical detection of a current-induced spin chemical potential difference on the surface of a $SmB_6$ single crystal. We clearly observe a surface-dominated spin voltage, which is proportional to the projection of the spin polarization onto the contact magnetization, is determined by the direction and magnitude of the charge current and is strongly temperature-dependent due to the crossover from surface to bulk conduction. We estimate the lower bound of the surface state net spin polarization as 15% based on the quantum transport model providing direct evidence that $SmB_6$ supports metallic spin helical surface states.

**INTRODUCTION**

Recent theoretical study identifies $SmB_6$ as a member of a newly classified family of strong topological insulators, topological Kondo insulators[1, 2, 3, 4], in which topologically protected surface states reside in the bulk Kondo band gap at low temperatures due to strong spin–orbit coupling. Following the measurement of the robust surface conduction below several Kelvin[5, 6, 7] superimposed with bulk insulating behaviour with a *d–f* hybridization induced gap in the range of 10–20 meV at intermediate temperatures below 300 K, many experimental efforts have been designed to probe the topological nature of the surface conducting states in this material[8, 9, 10, 11, 12, 13].

While the two-dimensional nature of the surface band and surface spin polarization has been confirmed by surface-sensitive probes such as photoemission[11, 14], and, most recently, the magnetic resonance induced spin pumping technique showed the possibility to inject a non-equilibrium spin current into the surface of $SmB_6$[12, 13], direct electrical measurement of surface current-induced spin–momentum locking, the unique feature of topological insulators, has not been performed in a simple transport geometry.

**MATERIALS AND METHODS**

**Material growth.** Single crystals of $SmB_6$ were grown with Al flux, starting from elemental Sm and B with the stoichiometry of 1 to 6 in a ratio of $SmB_6$ : Al = 1 : 200–250. The initial materials were placed in an alumina crucible and loaded in a tube furnace under Ar atmosphere. The assembly was heated to 1250–1400 °C and maintained at that temperature for 70–120 hours, then cooled at −2 °C/hr to 600–900 °C, followed by faster cooling. The $SmB_6$ samples were put into sodium hydroxide to remove the residual Al flux.

**Device fabrication.** An Al layer of 2 nm was deposited on the polished (100) surface of $SmB_6$ by using electron beam evaporation followed by oxidizing on a hotplate in ambient conditions. The resulting thin Al oxide layer prevents direct contact of the ferromagnetic electrode with $SmB_6$. Standard e-beam lithography was used to make electrode patterns. A permalloy (Py) layer was used as a ferromagnetic detector for spin chemical potential measurement, with the lateral size of 150 x 150 $\mu m^2$ and thickness of 20 nm deposited and capped with 15 nm of Au using electron beam evaporation. Non-ferromagnetic contacts used for the source, drain and reference electrodes were formed by e-beam lithography patterning and Al oxide etching with a buffered oxide etchant followed by depositing Ti at 5 nm/Au 80 nm using electron beam evaporation. For the Au electrode acting as the wire bonding pad for the ferromagnetic contact, additional insulating layer was made below the metal layer by an overdosing electron beam on electron beam resist (PMMA 950A6) with a dose of 10000 $\mu C/cm^2$ (see the inset to Fig. 1c).

**Transport measurements.** The device was placed in a Quantum Design PPMS variable temperature cryostat for low-temperature electrical measurements. For current-induced spin polarization measurement, both DC and AC type four-point probe measurements were performed. A DC (AC) current was applied through the $SmB_6$ surface channel from the non-magnetic contact source to drain using Keithley 2400 (Keithley 6221) instruments, and a Keithley 2182 nanovoltmeter (Stanford Research Systems SR830 Lock-in amplifier) was used for detecting a voltage difference between the Py and reference Au contact.

## RESULTS

**Principle of potentiometric spin measurement.** Figure 1a shows the simplified spin–momentum relation in SmB$_6$ near the Fermi energy revealed by recent photoemission studies[11] for both $\bar{X}$ and $\bar{\Gamma}$ high-symmetry points. When the electrons are placed under an electrochemical potential difference, for example, in the x-direction, the electrons moving to the right (red arrow) have higher occupation than the electrons going to the left (blue arrow), which leads to a difference between the electrochemical potentials for spin up $\mu_\uparrow$ and down $\mu_\downarrow$. This momentum asymmetry leads to a spin polarized current in the y-direction due to spin–momentum locking, thus the presence of the spin–momentum locking property in the SmB$_6$ surface state can be shown electrically by detecting the spin polarized current generated by the electrochemical potential difference. We note that the surface bands in SmB$_6$ exhibit spin–momentum relation, hence the sign of the expected spin voltage[15, 16, 17], opposite to other topological insulators like Bi$_2$Se$_3$[4, 18], as we discuss below.

Here, we use a specially designed potentiometric geometry, as shown in Fig. 1c, to probe the aforementioned spin-dependent chemical potential difference induced by momentum imbalance. A bias current flows through a non-magnetic contact on the SmB$_6$ surface with the (100) crystallographic plane in the x-direction, and the transverse voltage $V_{xy}$ is measured between a permalloy magnetic contact (Py) and a reference non-magnetic contact. Figure 1d shows the electrochemical potential for spin up $\mu_\uparrow$ and down electrons $\mu_\downarrow$ as a function of the SmB$_6$ channel position. The ferromagnetic contact can detect spin-dependent electrochemical potentials according to its magnetization direction, and the spin voltage corresponding to the difference between the electrochemical potentials for spin up and down can be expressed as follows, based on the quantum transport model[19, 20].

$$\Delta V_{xy} = V_{xy}(M) - V_{xy}(-M) = |I_b| R_B P_{FM} (p \cdot M_u), \qquad (1)$$

where $\Delta V_{xy}$ is the spin voltage defined as the difference of the measured electrochemical potential $V_{xy}$ between the opposite detector magnetization $M$ controlled by the external magnetic field $\vec{H}_{ext}$, and $\Delta V_{xy}$ is proportional to the magnitude of the bias current $|I_b|$, ballistic resistance of the channel $R_B$, spin polarization of the ferromagnetic detector $P_{FM}$, and inner product between the spin polarization of the TI surface $p$ and unit vector $M_u$ along the magnetization of the ferromagnetic detector. As Eqn. 1 indicates, we measure $\Delta V_{xy}$ as a function of the experimental parameters such as the direction and magnitude of $I_b$, $M$ and temperature $T$ and confirm the current-induced spin polarization on the surface of $SmB_6$.

**Measurement of the current-induced spin polarization.** We first show that an electrochemical potential bias can induce a spin polarization that is reflected in a non-zero $\Delta V_{xy}$ measured by the magnetic detector. As shown in Fig. 2a, $V_{xy}$ is measured by sweeping an external magnetic field in the y-axis $H_y$ to control the detector magnetization direction while applying $I_b$ along the x-axis. Figures 2b and 2c show representative spin voltage data recorded with $I_b$ of +100 μA and −100 μA, respectively. As the current is applied in the +x (−x) direction, where the momenta of the electrons are in the −x (+x) direction, the direction of the current-induced spin polarization is parallel to the −y (+y) axis due to the anti-clockwise spin texture (see the insets of Fig. 2b and 2c for the detailed directions of the electron momentum, $p$, and $M$). In both Fig. 2b and 2c, a high (low) voltage is measured when the $M$ of the ferromagnet is parallel (antiparallel) to $p$[15, 21] (see Supplementary Information S1), consistent with the spin–momentum relation in $SmB_6$[2, 4]. The measured voltage switches near the coercive field of Py (see Supplementary Information S2 and S11), which can be explained by the fact that the sign of $\boldsymbol{p}\cdot\boldsymbol{M_u}$ in Eqn. 1 changes when the direction of the magnetization is switched. Moreover, the polarity of the hysteresis loop in Fig. 2b and 2c is the opposite reflecting the fact that the current-induced spin polarization direction is dependent on the direction of $I_b$. More specifically, the measured $\Delta V_{xy}$ is shown in Fig. 2d as a function of $I_b$ and exhibits a clearly linear response. Therefore, the measured ferromagnetic spin voltage as a function of the magnitude and direction of $M$ and $I_b$ strongly indicates electrical measurement of the current-induced spin polarization on the surface of $SmB_6$. Additionally, we find spin-to-charge conversion in the $SmB_6$ surface state through a reciprocal geometry measurement consistent with the Onsager reciprocal relation (see Supplementary Information S3). In Supplementary Information S4, we further discuss the degradation of $\Delta V_{xy}$ in the non-linear transport regime at high $|I_b|$ due to Joule heating and subsequent bulk carrier population [22].

The second equality in Eqn. 1 provides a quantitative estimation of the degree of surface ensemble spin polarization $|p|$. We estimate $|p|$, extracted from the slope of $\Delta V_{xy}(I_b)$ dependence (see Fig. 2d), as ~ 15%, based on the following experimental conditions and assumptions: $1/R_B$ is given by $q^2/h$ times the number of mod6es $k_F W/\pi$, where $W = 500$ μm is the width of the current channel, the total Fermi wave number $k_F$ (0.218 Å$^{-1}$) [23] can be determined as $k_{F\alpha}+2k_{F\beta}$, where $k_{F\alpha}$ and $k_{F\beta}$ are the Fermi wave numbers of the α and β bands in the first Brillouin zone, respectively, and the $P_{FM}$ of Py at low temperatures is 0.2 [24]. We note that this is a conservative estimation of $|p|$ since we assume 100% single-surface-dominated conduction, as well as perfect ferromagnetic detector efficiency. The inclusion of experimental imperfections, such as the current path through not only the top but also the bottom surface[6], possible (although small) current leakage through bulk or imperfect detection efficiency of the ferromagnetic detector, will only make the estimation

of |*p*| higher, so that our estimation sets the lower bound of the surface current-induced ensemble spin polarization of SmB$_6$.

**Exclusion of possible artifacts.** To further confirm the origin of $V_{xy}$, in particular to exclude the possibility that the hysteresis loops of $V_{xy}$ in Fig. 2 could be due to spurious effects such as the planar Hall effect from the fringe field of the ferromagnetic detector[25], we perform a control experiment by applying $\vec{H}_{ext}$ in the x-direction, where *M* is orthogonal to *p*. Figures 3a and 3b show schematics of the $V_{xy}$ measurement configurations when applying $\vec{H}_{ext}$ in the y-direction and x-direction, respectively, and the corresponding measurement results are shown in Fig. 3c and 3d. Compared to Fig. 3c, when the magnetic field is swept in the x-direction, we do not observe the spin chemical potential difference $\Delta V_{xy}$ at high positive or negative $H_x$, reflecting that the measured $\Delta V_{xy}$ clearly follows the current-induced spin polarization origin, as *p*·*M*$_u$ term in the Eqn. 1 indicates. The intermittent non-zero signal in Fig. 3d likely stems from the magnetic domain, whose transient magnetization direction has some y-axis component. The result shows that the measured $\Delta V_{xy}$ depends on the projection of the spin polarization onto the detector magnetization direction consistent with the spin-texture model of SmB$_6$.

**Temperature dependence of the spin voltage.** We now turn to discussing the surface origin of the measured spin voltage. The potentiometric measurement performed at 1.8 K already shows evidence of spin polarization in the surface-dominated transport regime. We further confirm this by investigating the temperature dependence of $\Delta V_{xy}$ with the concurrently measured temperature-dependent charge conduction. As shown in Fig. 4a, the temperature-dependent electrical resistance *R*(T) of SmB$_6$ exhibits thermally activated behaviour at intermediate temperatures below 12 K, before saturating at an approximately temperature-independent value below several Kelvin, typically 4 K, strongly supporting the model of the insulating bulk with metallic surface states, as previously probed by other techniques[5, 7]. Performed at temperatures ranging from 1.8 to 4.5 K (marked by red dots in Fig. 4a), Fig. 4b–4g show $V_{xy}$ as a function of $H_y$ under $|I_b|$ of 100 μA. Strong temperature dependence is observed with vanishing $\Delta V_{xy}$ at ~ 4 K, which closely follows the crossover from surface- to bulk-dominated charge conduction around the same temperature (see the inset to Fig. 4a). Moreover, when we consider parallel two channels combined with surface channel resistance independent of temperature and bulk channel resistance dependent on temperature, the ratio between current flowing through surface channel at different temperature is determined by the ratio between total resistance at different temperature. From 1.8 to 4 K, the overall resistance is reduced by 60%, while $\Delta V_{xy}$ nearly completely vanishes, which indicates that the additional spin polarization reduction such as spin flip scattering between the surface and bulk conduction channels or spin current cancellation between opposite spin polarization of the surface and bulk spin Hall effect may be important to understand net spin polarization at elevated temperatures. Overall, the results not only confirm that the measured $\Delta V_{xy}$ indeed originates from a surface-dominant effect, but also show that bulk $SmB_6$ does not exhibit spin–momentum locking.

**Magnetic field angle dependence of the spin voltage.** The fact that the surface-dominated $\Delta V_{xy}$ shows a clear in-plane anisotropy with respect to the directions of $I_b$ and $M$ provides strong evidence for a spin–momentum locked surface spin polarization in $SmB_6$. However, the described measurements alone do not distinguish the in-plane vs. out-of-plane nature of the spin polarization in $SmB_6$. We finally discuss this by showing an angle-resolved spin voltage measurement. We apply 2 T of $|\vec{H}_{ext}|$ to ensure saturation of $M$ to the direction of $\vec{H}_{ext}$ and rotate the field in the y-z plane as shown in Fig. 5a, where γ is the angle between the direction of $M$ and the y-axis. $|I_b|$ of 300 μA is applied in the +x or −x direction, and γ-dependent $V_{xy}$ is recorded at 2 K. For an accurate spin voltage analysis, the Hall voltage with sin γ dependence (proportional to the z-component of an external magnetic field), as well as higher-order magneto-resistance components were excluded from the raw $V_{xy}$ data (see Supplementary Information S5). Figure 5b shows the resulting $\Delta V_{xy}(\gamma)$ in polar coordinates normalized to the maximum spin voltage $\Delta V_{xy,\,max}$. The vanishing spin voltage in the out-of-plane configuration ($\gamma = 90°$ or $270°$), while $\Delta V_{xy,\,max}$ occurs near $\gamma = 0°$ and $180°$, clearly indicates an overall in-plane ensemble spin polarization on the surface of $SmB_6$ within the experimental error of the field angle calibration of several degrees. We note that the actual angular distribution of $\Delta V_{xy}$ in $SmB_6$ may have a richer structure than a simple cosine function (compare Fig. 5b, the data and the fit represented by the solid line), which may stem from the non-atomically flat surface morphology of the polished $SmB_6$ crystal or, possibly, from the combined effects of multiple conduction surface bands in $SmB_6$. However, more precise determination of $\Delta V_{xy}(\gamma)$ is not possible with the signal-to-noise ratio of the current experiment, and we leave it for the future work.

**DISCUSSION**

A simple potentiometric geometry with a ferromagnetic contact enables direct electrical measurement of spin chemical potential in the proposed topological Kondo insulator $SmB_6$. Unlike the situation in conventional topological insulators like $Bi_2Se_3$ [26, 27], the location of the Fermi energy, pinned near the hybridization-induced gap due to the Kondo mechanism, guarantees surface-dominated transport in $SmB_6$ at low temperatures, thereby allowing clear surface spin voltage measurement even without extrinsic chemical doping[28] or gating technique[29, 30] conventionally used for non-ideal topological insulators[27]. The absence of chemical doping or surface gating makes it very likely that the $SmB_6$ studied here is free from surface band bending related two-dimensional electron gas[31], further confirming that the measured spin voltage mainly stems from the intrinsic surface spin polarization. However, we do not rule out, although it is estimated to be small[19], the possible contribution from the detailed spin textures of the structural-symmetry-broken Rashba surface states, the spin voltage sign of which cannot be distinguished from the topologically protected surface state in the case of $SmB_6$, as noted earlier, since, unlike $Bi_2Se_3$, $SmB_6$ shows the same anti-clockwise spin texture as that of Rashba surface states in most cases[11, 19]. Further systematic study on the in-plane magnetization angle dependence or high-resolution spin-resolved photoemission combined with theoretical calculation is needed to fully separate the topologically nontrivial and trivial spin polarization contributions. Nevertheless, with the ability to clearly measure the intrinsic, surface-dominated spin polarization in strongly correlated systems, this approach provides potential for both fundamental and applied spin transport studies in newly proposed topologically nontrivial states of matter [32].

**CONFLICT OF INTEREST**

The authors declare no competing financial interests.

**ACKNOWLEDGMENT**

This research was supported by the Basic Science Research Program (Grant No. NRF-2015R1C1A1A02037430) and Priority Research Centers Program (No. 2015R1A5A1037668) through the National Research Foundation of Korea (NRF) funded by the Ministry of Science, ICT and Future Planning. Preparation of $SmB_6$ material was funded by AFOSR (FA9550-14-1-0332) and by the Gordon and Betty Moore Foundations EPiQS Initiative through Grant GBMF4419. Electrical measurements used shared facilities funded by the KIST institutional program and the National Research Council of Science&Technology (NST) grant (No. CAP-16-01-KIST).

Supplementary information is available at the NPG Asia Materials website

(http://www.nature.com/am)

**Figure captions**

**Figure 1** Surface spin texture and potentiometric spin measurement in $SmB_6$. **(a)** Anti-clockwise spin texture for the surface band in $SmB_6$ near the Fermi energy shown by spin and angle resolved photoemission spectroscopy and first-principles calculations[2, 11]. **(b)** Degree of electrons

occupation in channel and spin-dependent electrochemical potential under electrochemical bias. The length of the red and blue arrows indicates the degree of moving electrons with spin *S*. **(c)** Measurement configuration for detecting spin voltage. Inset: optical microscope image of the device. The scale bar is 100 μm. **(d)** Electrochemical potential with respect to the position of the channel, illustrating concepts of the potentiometric spin measurement.

**Figure 2** Measurement of current-induced surface channel spin polarization. **(a)** Schematic of the transport measurements. A constant current $I_b$ is applied in the x-direction while sweeping a magnetic field in the y-direction, which is orthogonal to the bias current direction (co-linear with the spin polarization direction). **(b)**, **(c)** Magnetic field dependence of the voltage measured at the ferromagnetic contact for $I_b$ of **(b)**, +100 μA and **(c)**, −100 μA. **(d)** Dependence of spin voltage $\Delta V_{xy}$ on $I_b$ measured at 1.8 K.

**Figure 3** Magnetization orientation dependence of spin voltage. **(a), (b)** Measurement configuration of spin voltage $\Delta V_{xy}$ with detector magnetization aligned **(a),** in the y-direction (perpendicular to current) and **(b),** in the x-direction (parallel to current). **(c), (d)** $V_{xy}$ as a function of an external magnetic field swept **(c),** in the y-direction, and **(d),** in the x-direction.

**Figure 4** Temperature dependence of the spin voltage and resistance of SmB$_6$. **(a)** Electrical resistance of SmB$_6$ as a function of temperature. Inset: spin voltage $\Delta V_{xy}$ as a function of temperature. The size of the error bar is smaller than the blue dot size. **(b)-(g)** $V_{xy}$ measured by sweeping $\vec{H}_{ext}$ parallel to the y-axis under a bias current of 100 μA at different temperatures, 1.8 K **(b)**, 2.5 K **(c)**, 3 K **(d)**, 3.5 K **(e)**, 4 K **(f)** and 4.5 K **(g)**.

**Figure 5** Angle-resolved current-induced spin polarization. **(a)** Schematic measurement configuration for potentiometric measurement with rotating magnetic field in the y-z plane, where γ is angle between the applied external magnetic field and the y-axis. **(b)** Polar plot of normalized $\Delta V_{xy}$ as a function of γ showing the in-plane character of the current-induced spin polarization in $SmB_6$.

Figure 1

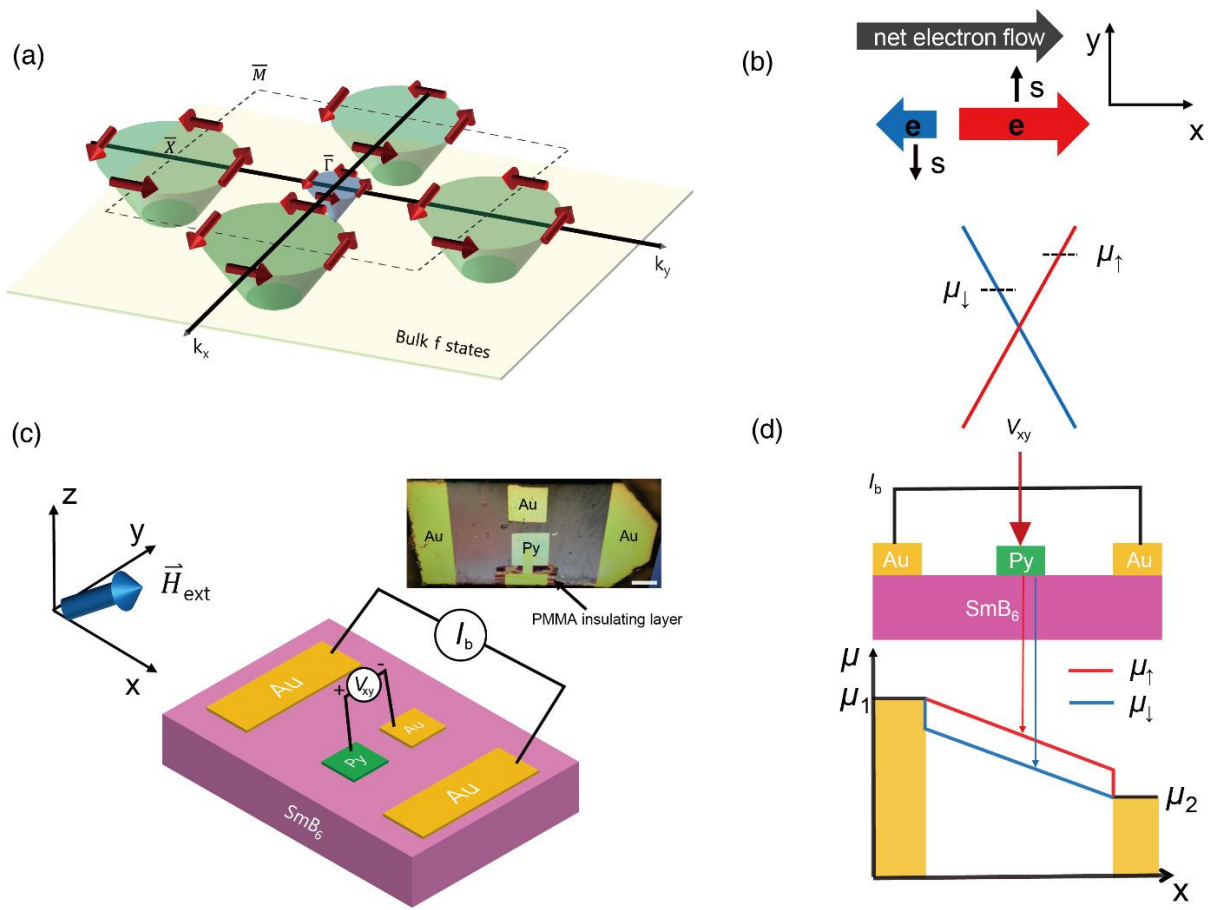

Figure 2

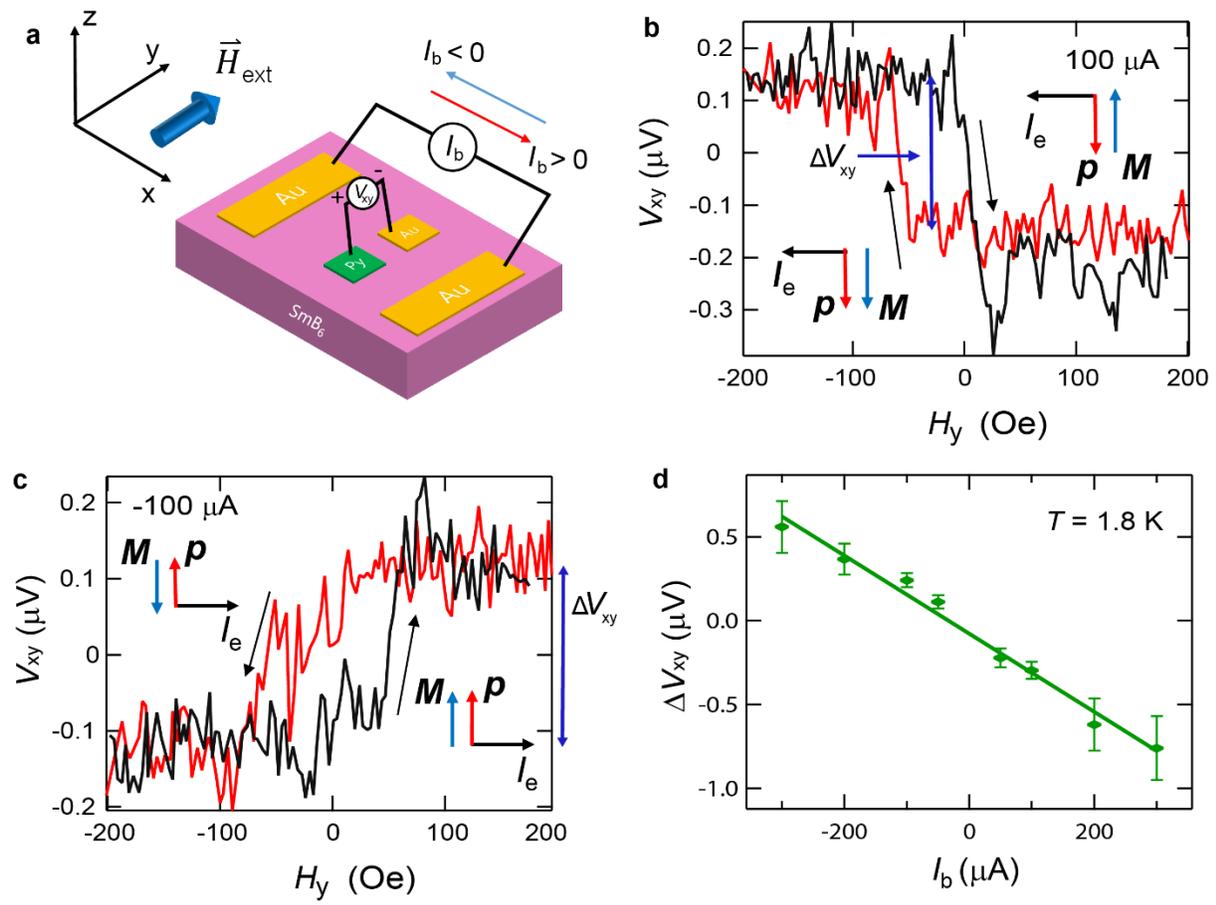

Figure 3

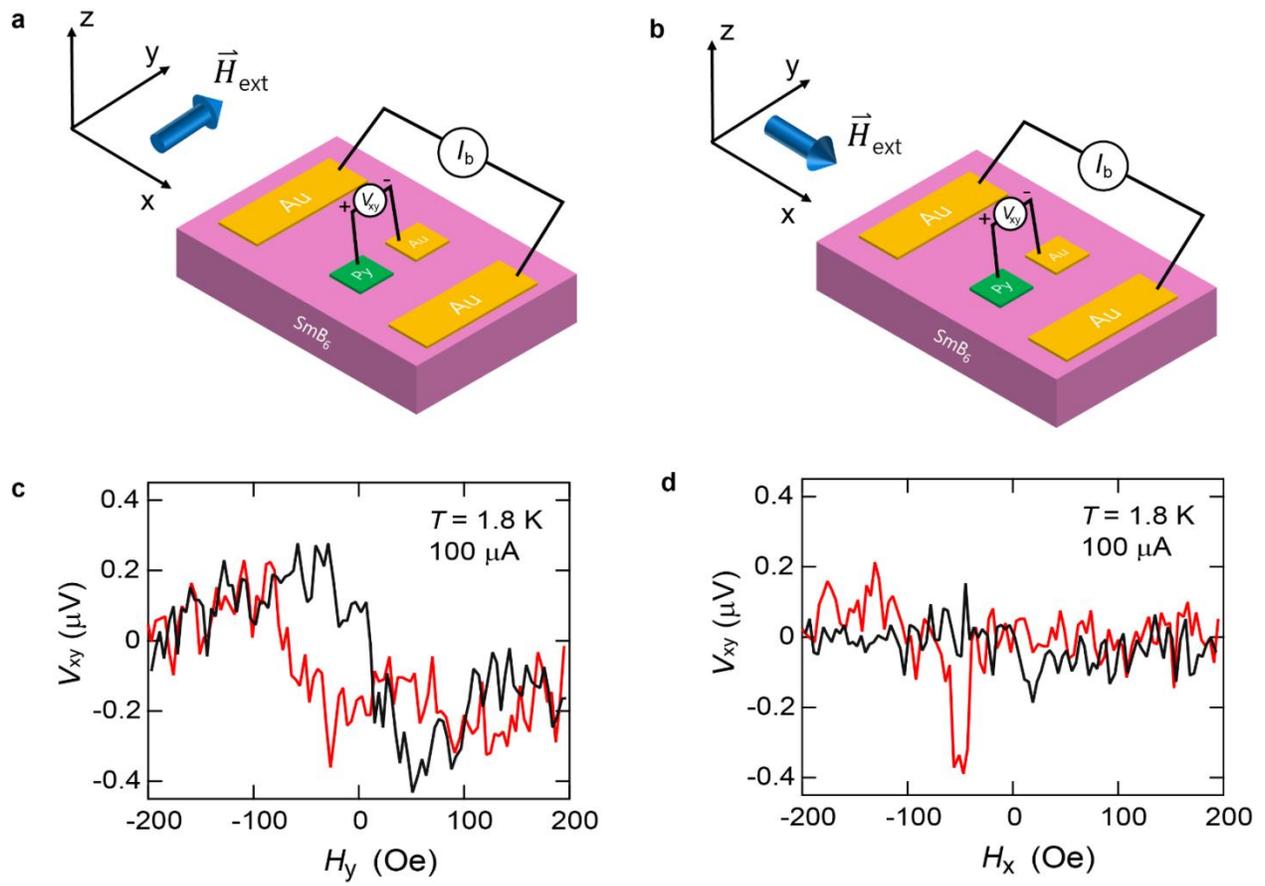

Figure 4

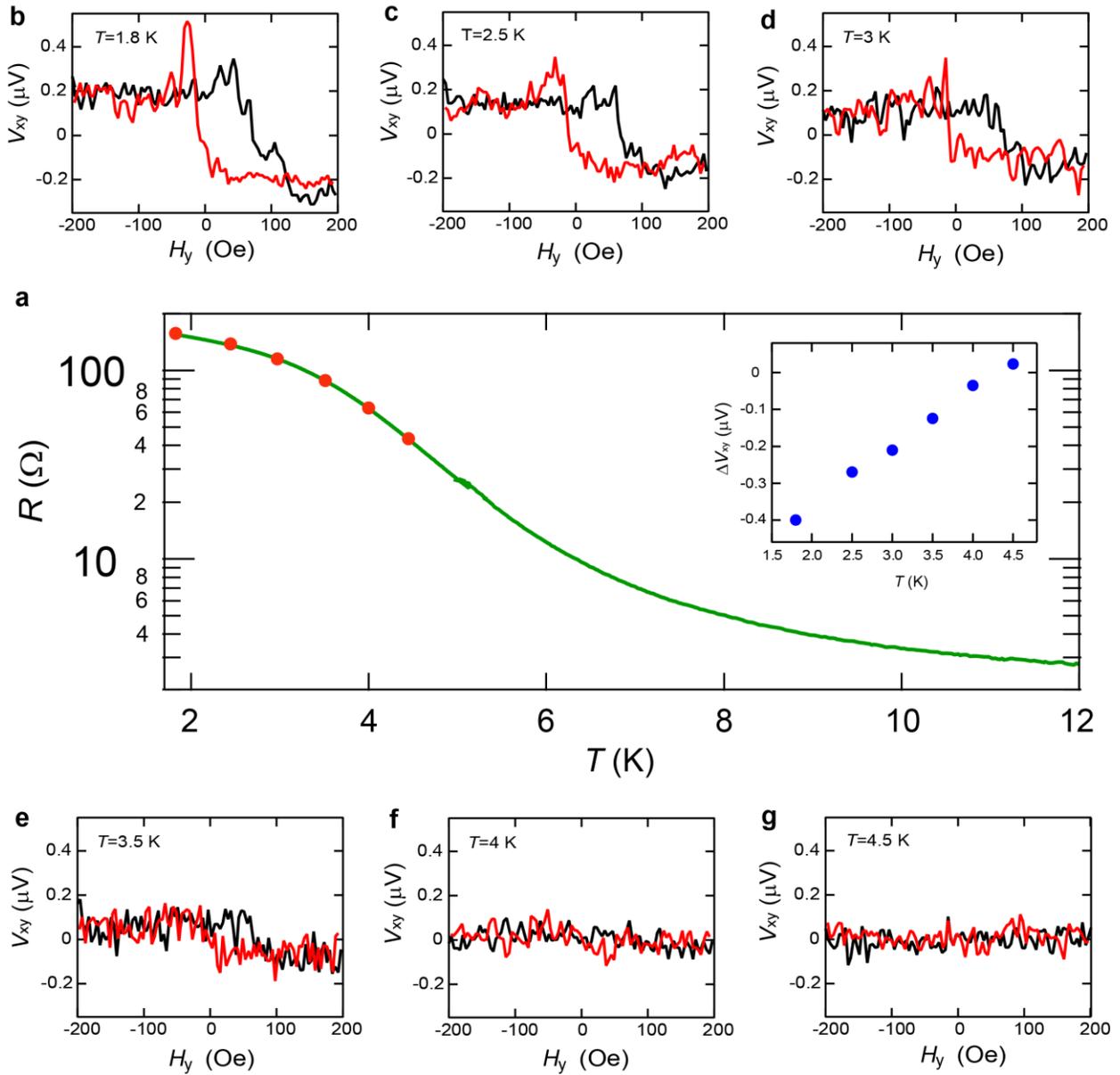

Figure 5

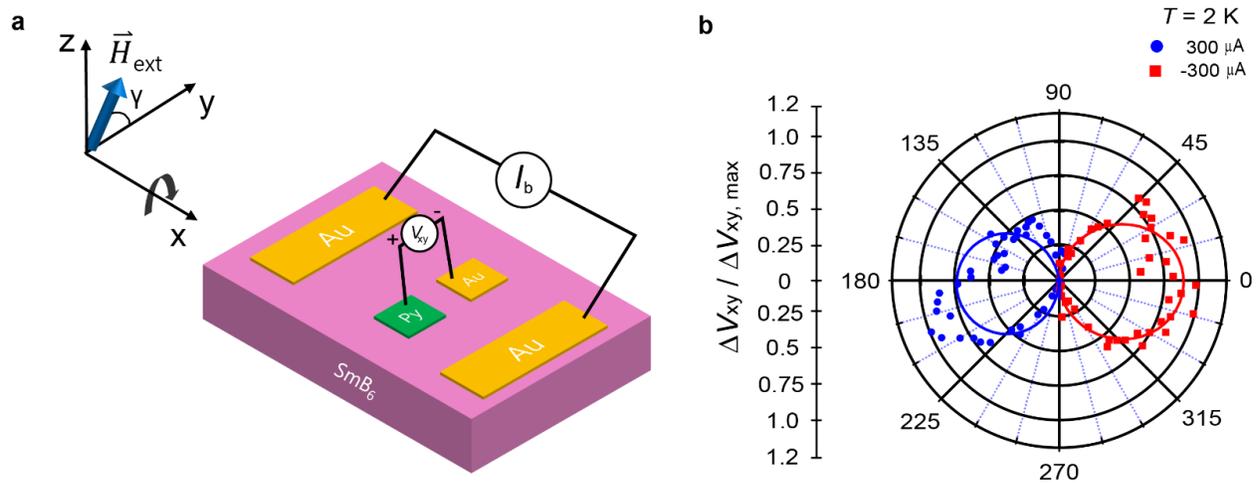